\documentstyle[psfig]{caosp}
\begin{document}
\pubyear{1993}
\volume{23}
\firstpage{7}
\htitle{The evolutionary state of metallic A-F giants}
\hauthor{P. North {\it et al.}}
\title{The evolutionary state and fundamental parameters of metallic A-F giants}
\author{P. North \and D. Erspamer \and M. K\"unzli}
\institute{Institut d'Astronomie de l'Universit\'e de Lausanne,\\
CH-1290 Chavannes-des-Bois, Switzerland}

\date{\today}
\maketitle
\begin{abstract}
Using Hipparcos parallaxes, we show that the metallic A-F giants
found by Hauck (1986) on the basis of their high $\Delta m_2$ index
in Geneva photometry are on average more evolved than their non-metallic
counterparts. Their mass distribution, rate of binaries and $v\sin i$
are shown to be incompatible with those of Am stars, so that they cannot be
descendants of the latter. They might be former normal stars going through
a short metal-rich phase at the very end of their life on the Main Sequence.
\end{abstract}
\section{Introduction}
The metallic A-F giants, found by Hauck (1986) on the basis of their high
$\Delta m_2$ index, have an abundance pattern similar to that of Am stars,
except for Ca and Sc which have a more or less solar abundance (Berthet 1990,
1991). Their chemical anomalies closely resemble those of the $\delta Del$
stars, which seem to be evolved Am stars (Kurtz 1976). Therefore, it appeared
natural to consider the metallic A-F giants as evolved Am stars too. Indeed,
the theory of radiative diffusion foresees that calcium, which had sinked to
large atmospheric depths in the beginning of the star's life would be
finally dredged up by the increasingly deeper outer convective zone as the
star would reach the giant phase. This scenario was advocated by Berthet (1992).

Hereafter, we reconsider this question from a different standpoint
and examine the evolutionary state of the metallic A-F giants in the light
of the Hipparcos results.

\section{Rotational velocity and rate of binaries}
If metallic A-F giants were indeed evolved Am stars, they should share with
them two essential characteristics: slow rotation and high rate of binaries.
Since giants have a larger radius, they should rotate even more slowly than
Am stars, merely by conservation of angular momentum. However, a glance at the
distributions of the $v\sin i$ values of Am, metallic and non-metallic A-F
giants suffices to cast serious doubts on the idea of an evolutionary link
between Am and metallic giants (see Figure 11 of K\"unzli \& North 1997):
$v\sin i$ is uniformly distributed
between 0 and 150~km\,s$^{-1}$ for metallic giants, while it shows a maximum
at 30-40~km\,s$^{-1}$ and remains smaller than 100 km\,s$^{-1}$ for Am stars.
The rate of binaries has been examined using both data from the literature
and observations done at Observatoire de Haute-Provence (CNRS), France, with
the Aur\'elie spectrograph attached to the 1.52m telescope in 1994.
For metallic A-F giants, the  rate of binaries with orbital
periods shorter than 1000 days is no more than 
about 23\%, matching very well the figure (21.7\%) found by Duquennoy \& Mayor
(1991) for G dwarfs. This figure is to be compared with the result of
Abt \& Levy (1985), who found 75\% of binaries among Am stars. Here again, the metallic giants lack an essential characteristic of Am stars,
namely a high rate of binaries (K\"unzli \& North 1997).

If metallic A-F giants are not evolved Am stars, what are they? the simplest
alternative scenario is that every A star goes through a short ``metallic'' 
phase at the end of its life on (or just beyond) the Main Sequence.

\section{Evolutionary state and fundamental parameters}
Practically all stars considered by Hauck (1986) have been measured by
Hipparcos, and many of them are closer than 100 pc or have, in any case, a
relative precision on the parallax better than $\sim 20\%$.
The resulting HR diagram shows that the metallic giants
tend to be more evolved on average than their normal counterparts, as confirmed
by the cumulative distributions of the $\log g$ values.
The $\log g$ values were obtained from the luminosities deduced from the
Hipparcos parallaxes, from the $T_{\mathrm eff}$ deduced from Geneva photometry
calibrated by K\"unzli et al. (1997) and from masses interpolated in the
evolutionary tracks of Schaller et al. (1992). The KS test shows a very
significant difference between the distributions of $\log g$ for metallic
and normal giants. The mass distribution of the metallic giants
is strongly peaked at 2 M$_\odot$, while the mass distribution of Am stars
is peaked at 1.5 M$_\odot$ (North 1993). This difference alone does not
completely exclude, however, any possible link between Am stars and metallic 
giants, because low mass Am stars would have evolved into giants with a
$T_{\mathrm eff}$ cooler than the limit above which radiative diffusion still
works.

\section{Conclusions}
There is a set of rather compelling arguments against the idea that
metallic A-F giants would be evolved Am stars. On the other hand, the precise
Hipparcos parallaxes allow for the first time to pinpoint these metallic
giants in the HR diagram and to show that they all have $\log g \leq 3.8$.
This seems to add some weight to the alternative idea that these giants have
nothing to do with Am stars but may be former ``normal'' A stars going through
a short phase where, for some as yet unclear reason, radiative diffusion is
allowed to enhance the metallic abundance in their atmosphere.
\begin{figure}[hbt]
\psfig{figure=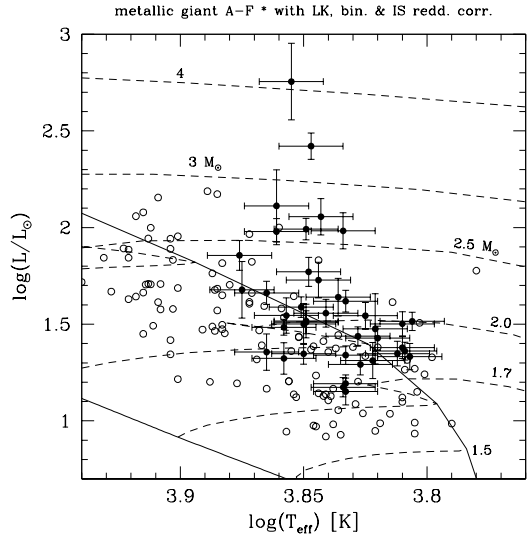,height=7.8cm}
\caption{HR diagram of all giants considered by Hauck (1986). Full dots:
$\Delta m_2 > 0.013$, open dots: $\Delta m_2\leq 0.013$. ZAMS, TAMS and 
evolutionary tracks for $Z=0.020$ are from Schaller et al. (1992).}
\label{fp}
\end{figure}

\end{document}